\documentstyle[preprint,aps,eqsecnum]{revtex}
\newcommand{\be}{\begin{eqnarray}}
\newcommand{\e}{\end{eqnarray}}

\begin{document}
\draft
\tighten

\preprint{DO-TH 02/18, MKPH-T-02-21}

\title
{Helicity-dependent Twist-two and Twist-three Generalized Parton 
Distributions in Light-Front QCD}

\author{{\bf A. Mukherjee$^{a}$}\thanks{e-mail: asmita@physik.uni-dortmund.de} 
{\bf and M. Vanderhaeghen$^{b}$}\thanks{e-mail: marcvdh@kph.uni-mainz.de}\\
 $^a$ Theoretische Physik IV, Universit\"at Dortmund, D 44221
Dortmund, Germany\\
 $^b$ Institut f\"ur Kernphysik, Universit\"at Mainz, D-55099, Mainz, Germany}

\date{February 16, 2003}

\maketitle
\begin{abstract}
We investigate the helicity dependent twist-two and twist-three 
generalized parton distributions in
light-front Hamiltonian QCD for a massive dressed quark target. Working in the
kinematical region $\xi< x <1$,  we obtain the splitting
functions for the evolutions of twist-two quark and gluon distributions in a
straightforward way. For the twist-three distribution, we find that
all contributions are proportional to the quark mass and thus
the twist-three distribution is directly related to the chiral symmetry
breaking dynamics in light-front QCD. We also show that the off-forward 
Wandzura-Wilczek relation is violated in perturbative QCD for a massive
dressed quark.We calculate the quark mass correction to the WW relation in the off-
forward case and show that it is related to $h_1(x)$ in the forward limit. We extract
 the 'genuine twist three' part of the matrix element in the forward case and verify 
the Burkhardt-Cottingham  and Efremov-Leader-Teryaev sum rules.
\end{abstract}
\vskip .2in
{\it Keywords: Generalized parton distributions, Light-front Hamiltonian, Perturbation theory}
\vskip .2in
PACS : 11.10.Ef, 12.38.Bx, 13.60.Hb,  11.15.Bt
\section{Introduction}
The generalized parton distributions (GPDs) \cite{dieter} are being studied 
intensively
in recent years. GPDs are  hybrids of the usual parton distributions measured
in inclusive processes like deep inelastic scattering (DIS) and form factors
measured in elastic exclusive processes. In general, they can be expressed
as {\it off-forward} matrix elements of light cone {\it bilocal} operators. In the
forward limit, GPDs reduce to normal parton distributions, which can then
be expressed as {\it forward} matrix elements of light-cone bilocal operators.
The moments of GPDs over the parton momentum fraction $x$ give form factors, 
which are given in terms
of off-forward matrix elements of {\it local} operators.  Thus, GPDs have a much 
richer structure
and they connect various processes, both inclusive and exclusive. They
provide new and important informations about the structure of the hadron.
They can be probed in deeply virtual Compton scattering (DVCS)  and hard
exclusive production of vector mesons (for recent reviews on GPDs and hard
exclusive reactions, see \cite{rev1,rev2,rev3} and references therein).  

The GPDs have been investigated recently in light-cone formalism by several
authors and an overlap representation of the plus component in terms of light-cone wave functions
has been given \cite{brod,kroll}.  GPDs have also been constructed using light-cone model
wave functions \cite{miller}. The transverse and the minus components are somewhat
complicated since they involve the constraint field $\psi_-$. They are
usually called 
'bad' components, since the operators in these cases involve interaction terms
and are higher twist objects. In other words, they involve direct
quark-gluon dynamics. To interpret the experimental results for DVCS on a
proton target in the presently accessible $Q^2$ range
\cite{hermes,clas,h1,zeus}, it is of primordial
importance to understand the effect of higher twist components. 
The perpendicular, or the twist-three component of the
operator has been investigated  using the Wandzura-Wilczek approximation
\cite{wan},
where the explicit interaction dependent parts of the operator as well as
the quark mass terms were neglected. In this case, the twist-three matrix
element can be expressed solely in terms of twist-two GPDs. 
In the forward limit, these relations reduce to the
Wandzura-Wilczek relation for the transversely polarized structure function
$g_T$ \cite{ww}. There is no theoretical justification for this
approximation. In the forward case, recent experimental results for the 
transverse polarized structure function indicate that the deviation from
the Wandzura-Wilczek approximated form is substantial in some kinematic
range for a nucleon target \cite{exp}.
Therefore, it is interesting to make a full calculation of the off-forward
twist-three matrix element, within the context of perturbative QCD, taking 
into account the explicit interaction
dependence of the operator, the mass  as well as the intrinsic transverse momentum of 
the partons. A convenient tool is based on the light-front Hamiltonian
description of composite systems utilizing many-body wave functions. Instead
of a hadron target, here we consider a simpler target like a quark
dressed with a gluon and calculate the off-forward matrix elements within
the context of perturbative QCD. The two-particle wave function is given in
terms of the light-front QCD Hamiltonian \cite{hari1}. This approach has been used
extensively in the recent years to calculate polarized and unpolarized
distribution functions in DIS, twist-two \cite{hari2,rajen}, twist-three
\cite{gt,hari3} and twist four \cite{hari4}, as well as the transversity 
distribution \cite{tran}, and the transverse momentum dependent
distributions \cite{metz}. Recently we have also
extended it to calculate the off-forward matrix elements of light-front 
bilocal vector operators \cite{vec}. We verified the helicity sum rule in perturbation
theory and showed the effect of quark mass in the twist-three matrix
element, which is absent in the forward limit.

In this work, we  investigate the helicity dependent generalized
distributions, which are the off-forward matrix elements of light-front axial
vector operators, in light-front Hamiltonian perturbation theory for a
dressed quark target. We restrict ourselves to the region $\xi<{\bar x}<1$, of the generalized distributions, 
where $\xi$ is the skewedness. In this region, the contributions come from the
overlaps of two-body wave functions upto $O(\alpha_s)$. We obtain the 
splitting functions corresponding to the
evolution of the twist-two helicity dependent distributions in a
straightforward way. In the twist-three distribution, we show the
contributions from the quark-gluon interaction dependent part, the quark mass
and the 
intrinsic transverse momentum dependent parts of the operator explicitly. We find that all the three
contributions are proportional to quark mass. We find that the
Wandzura-Wilczek (WW) relation is not satisfied in perturbative QCD in the
off-forward case for a massive quark, analogously as its forward counterpart. Our results also
 show that the twist-three distribution is
directly related to the dynamical effect of chiral symmetry breaking in
light-front QCD.  We calculate the mass correction to the WW relation in the off-forward case. This contribution is related to $h_1(x)$ in the forward limit. Using the quark mass correction to the WW relation, we also obtain the 'genuine twist three' part of the matrix element. We show that the first and second moments of this are zero which give the Burkhardt-Cottingham 
(BC) \cite{bc} 
and Efremov-Leader-Teryaev (ELT) \cite{elt} sum rules respectively.

The plan of the paper is as follows. In section II, we investigate the twist
two helicity dependent distributions, involving both the quark and the gluon
operators, for a dressed quark state. In section III, we calculate
the off-forward 
matrix element of the transverse component of the axial vector current. We investigate the WW relation in the 
off-forward case and the quark mass correction to it in section IV. Summary and
discussions are given in section V. The light front spinors for longitudinally and transversely polarized quarks 
are given in Appendix A. An outline of the derivation of the quark mass
term in WW approximation is given in Appendix B.
\section{Helicity Dependent Twist-two Distributions}
\subsection{Quark Distribution}
\vskip .2in
The twist-two helicity-dependent generalized distribution is given by
\be
\tilde F^{+}_{\lambda \lambda'}=\int {dz^-\over {8 \pi}} e^{{i\over 2}{\bar x}{\bar P}^+{z^-}}\langle P'
\lambda' \mid \bar \psi (-{ z^-\over 2}) \gamma^+ \gamma^5\psi ({ z^-\over 2}) 
\mid P \lambda \rangle.
\label{eq1}
\e
\vspace{0.2cm}
Here, $P,P'$ are the 4-momentums and $\lambda, \lambda'$ are the 
helicities of the initial and final states respectively. 

We work in the so called symmetric frame \cite{brod,kroll}. The momentum of the initial state
is $P^\mu$ and that of the final state is $P'^\mu$. The average momentum
between the initial and final state is then ${\bar P}^\mu={P^\mu+P'^\mu\over 2}$.
The momentum transfer is given by $\Delta^\mu=P'^\mu-P^\mu$, $ P'_\perp =
-P_\perp
={\Delta_\perp\over 2}$, skewedness $\xi=-{\Delta^+\over {2 {\bar P}^+}}$.
Without any loss of generality, we take $\xi >0$. We also get $\Delta^-={\xi
{\bar P}^2\over {\bar P}^+}$.

The above matrix element is conventionally parametrized in terms of the helicity 
dependent distributions, ${\tilde H(x,\xi,t)}$ and ${\tilde E}(x,\xi,t)$, where $t$ is 
the invariant momentum transfer. The matrix element can also be expressed in terms of 
overlaps of light-front wave functions. The operator is given by,
\be
\int {dz^-\over {8 \pi}} e^{{i\over 2}{\bar x}{\bar P}^+{z^-}}\bar \psi (-{ z^-\over 2}) 
\gamma^+ \gamma^5\psi ({ z^-\over 2})
={1\over {4 \pi}}\int dz^-  e^{{i\over 2}{\bar x}{\bar P}^+{z^-}}\psi^{+
\dagger} (-{ z^-\over 2}) \gamma^5 \psi^+ ({ z^-\over 2}).
\e
Here $\psi^+= \Lambda^+ \psi, \Lambda^\pm={1\over 2}\gamma^0 \gamma^\pm$.
The above expression is given in the light-front gauge,  
$A^+=0$, where the path-ordered exponential between the fermion fields in
the bilocal operator is unity. For simplicity we suppress the flavor indices.   
In the two-component representation, we have the dynamical fermion field,
\be
\psi^+(z)=\sum_{\lambda} \chi_\lambda  \int {dp^+ d^2p^\perp\over {2(2\pi)^3
{\sqrt p^+}}} [b(p,\lambda)e^{-iqz}+d^\dagger(p,\lambda)e^{ipz}],
\label{psi}
\e
and the dynamical gauge field,
\be
A^i(z)=\sum_{\lambda} \int {dq^+ d^2q^\perp\over {2(2\pi)^3 q^+}}[
\epsilon^i_\lambda a(q, \lambda)e^{-iqz}+h.c.],
\e
Here, $\chi_\lambda$ is the eigenstate of $\sigma_z$ in the two-component
spinor of $\psi^+$. We have used the light-front $\gamma$ matrix
representation:
\be
\gamma^0=\left (\begin{array}{cc}0&-i\\i&0\end{array}\right ),~~
\gamma^3=\left (\begin{array}{cc}0&i\\i&0\end{array}\right ),~~
\gamma^i=\left (\begin{array}{cc}-i {\tilde \sigma}^i&0\\0&i {\tilde \sigma}^i\end{array}\right
),
\e
with $\tilde \sigma^1=\sigma^2$ and $\tilde \sigma^2=-\sigma^1$.
$\epsilon^i(\lambda)$ is the polarization vector of the transverse gauge
field.

The Fock space expansion of the operator is given by,
\be
O^{+5}&=& 2 \sum_{s,s'}\int {dk^+ d^2k^\perp\over {2(2 \pi)^3 \sqrt k^+}}\int {dk'^+ 
d^2k'^\perp\over {2(2 \pi)^3 \sqrt k'^+}} \nonumber\\&&~~~~
\Big [ \delta(2 {\bar x} {\bar
P}^+-k'^+-k^+) b^\dagger (k,s)b(k',s')\nonumber\\&&~~~~~~~~~~~~~~~
+ \delta(2 {\bar x} {\bar
P}^++k'^++k^+) d (k,-s)d^\dagger (k',-s')
\nonumber\\&&~~ + \delta(2 {\bar x} {\bar P}^++k^+-k'^+) d (k,-s) b(k',s')
\nonumber\\&&~~~~~~~~~~~~~~~~ 
+ \delta(2 {\bar x} {\bar P}^++k'^+-k^+) b^\dagger(k,s)
d^\dagger(k',-s')\Big ] \chi^\dagger_{s} \sigma_3 \chi_{s'}.
\label{fock}
\e
We have, $k^+>0$,$k'^+>0$, $k^+-k'^+=p^+-p'^+=2 \xi {\bar p}^+$. In the
kinematical region, $\xi<{\bar x}<1$, only the first term in Eq.
(\ref{fock}) contributes
\cite{kroll}. We
restrict ourselves to this kinematical region.

We take the state 
$ \mid P, \sigma \rangle$ of momentum $P$ and helicity $\sigma$ to be a 
dressed quark
consisting of bare states of a quark and a quark plus 
a gluon:
\begin{eqnarray}
\mid P, \sigma \rangle && = \phi_1 b^\dagger(P,\sigma) \mid 0 \rangle
\nonumber \\  
&& + \sum_{\sigma_1,\lambda_2} \int 
{dk_1^+ d^2k_1^\perp \over \sqrt{2 (2 \pi)^3 k_1^+}}  
\int 
{dk_2^+ d^2k_2^\perp \over \sqrt{2 (2 \pi)^3 k_2^+}}  
\sqrt{2 (2 \pi)^3 P^+} \delta^3(P-k_1-k_2) \nonumber \\
&& ~~~~~\phi_2(P,\sigma \mid k_1, \sigma_1; k_2 , \lambda_2) b^\dagger(k_1,
\sigma_1) a^\dagger(k_2, \lambda_2) \mid 0 \rangle. 
\label{eq2}
\end{eqnarray} 
Here $a^\dagger$ and $b^\dagger$ are bare gluon and quark
creation operators respectively and $\phi_1$ and $\phi_2$ are the
multiparton wave functions. They are the probability amplitudes to find one
bare quark and one quark plus gluon inside the dressed quark state
respectively. Up to one loop, if one considers all kinematical
regions, there will be non-vanishing contributions from the overlap of
3-particle and one particle sectors of the state, this situation is similar
to QED \cite{brod}. In the kinematical region 
we are considering, such kind of overlaps are absent and it is sufficient
to consider dressing only by a single gluon. The state is normalized as,
\be
\langle P',\lambda'\mid P,\lambda \rangle = 2(2\pi)^3
P^+\delta_{\lambda,\lambda'} \delta(P^+-{P'}^+)\delta^2(P^\perp-P'^\perp).
\e
$\phi_1$ actually gives the normalization constant of the state \cite{hari2}:
\be
{\mid \phi_1 \mid}^2=1-{\alpha_s\over {2 \pi}} C_f
\int_0^{1-\epsilon} dx {{1+x^2}\over {1-x}}log
{Q^2\over \mu^2},
\label{c5nq}
\e
within order $\alpha_s$.  Here $\epsilon$ is a small cutoff on $x$.
 
The matrix element becomes,
\be
\tilde F^+&=& {\sqrt {1-\xi^2}}\Big [ \psi_1^* \psi_1 \delta(1-{\bar x})
\nonumber\\&& + \sum_{s_1,s_2,\lambda} \int d^2q_\perp
\psi_{2 s_1,\lambda}^{*\uparrow}({{\bar x}-\xi\over {1-\xi}},q^\perp+{1-{\bar
x}\over {1-\xi^2}}\Delta^\perp) \chi^\dagger_{s_1} \sigma^3 \chi_{s_2} \psi_{2 s_2 \lambda}^\uparrow 
({{\bar x}+\xi\over {1+\xi}}, q^\perp)\Big ].
\label{f+5}
\e
We have introduced Jacobi momenta $x_i$,${q_i}^\perp$ such that $\sum_i x_i=1$ and
$\sum_i {q_i}^\perp=0$ and the boost invariant wave functions,
\be
\psi_1=\phi_1, ~~~~~~~~~~~\psi_2(x_i,q_i^\perp)= {\sqrt P^+} \phi
(k_i^+,{k_i}^\perp).
\e
The first term in Eq. (\ref{f+5}) is the contribution from the single
particle sector and the second term is the contribution of the two-particle
sector of the state. Using light-front QCD Hamiltonian, the two-particle
wave function is given in terms of $\psi_1$ as : 
\be
\psi^{\sigma,a}_{2\sigma_1,\lambda}(x,q^\perp)&=& -{x(1-x)\over (q^\perp)^2}T^a 
{1\over {\sqrt {(1-x)}}} {g\over
{\sqrt {2(2\pi)^3}}} \chi^\dagger_{\sigma_1}\Big [ 2 {q^\perp\over
{1-x}}+{{\tilde \sigma^\perp}\cdot q^\perp\over x} {\tilde \sigma^\perp}
\nonumber\\&&~~~~~~~~~~~~~~~~~~
-i m{\tilde \sigma}^\perp {(1-x)\over x}\Big ]\chi_\sigma \epsilon^{\perp *}_\lambda \psi_1.
\label{psi2}
\e
where $g$ is the coupling constant, $ T^a $ is the usual $({1\over 2}$ of
Gell-Mann) color matrix  and $m$ is the bare mass of the quark. In
the denominator of the above expression, we have neglected terms of order
$m^2$ compared to $(q^\perp)^2$. 
Using Eq. (\ref{eq1}) and Eq. (\ref{psi2}), we see that the linear mass terms 
which cause helicity flip, are suppressed in the matrix element. The terms
quadratic in mass do not flip helicity, but they are suppressed too.
We calculate the helicity non-flip part.

Using Eq. (\ref{psi2}) we get,
\be
\tilde F^+={\sqrt {1-\xi^2}} \Big [ \delta(1-{\bar x})+{\alpha_s\over {2
\pi}}C_f log{Q^2\over \mu^2}
{(1+{\bar x}^2-2\xi^2)\over {(1-{\bar x})(1-\xi^2)}} \Big ] \psi_1^* \psi_1.
\label{plusres}
\e 
where $C_f={(N^2-1)\over 2N}$ for $SU(N)$.
We have cut off the transverse momentum integral at some scale $Q$ and
$\mu$ is the factorization scale separating hard and
soft dynamics \cite{gt}. Also, we have taken $\mid \Delta^\perp \mid $ to
be small. For convenience, we take $\Delta^2=0$. It is important to note that the entire
$\alpha_s$ dependency in Eq. (\ref{plusres}) comes from the state and the operator is independent
of interaction. The single particle matrix element receives a contribution
upto $O(\alpha_s)$ from the normalization of the state. Taking into account
the normalization contribution, we get \footnote{Here ${1\over (1-x)_+}$ is the usual (principal value) 
plus prescription.},
\be
\tilde F^+={\sqrt {1-\xi^2}} \Big [ \delta(1-{\bar x})+{\alpha_s\over {2
\pi}}C_f log{Q^2\over \mu^2}\Big ( {3\over 2} \delta(1-{\bar x}) +
{(1+{\bar x}^2-2\xi^2)\over {(1-{\bar x})_+(1-\xi^2)}}\Big ) \Big ].
\label{plusnor}
\e
The end point singularity at ${\bar x}=1$ is canceled by the contribution from 
the normalization of the state to the single particle matrix element,
similar to the helicity independent case \cite{vec}.
The splitting function can be easily extracted from the above expression :
\be
{\tilde P}_{qq}= C_f{1+{\bar x}^2-2\xi^2\over {(1-{\bar x})_+(1-\xi^2)}}.
\e
This agrees with the known result of \cite{ji} (when replacing 
 $\xi$ in Ref. \cite{ji} by $2 \xi$).

Turning next to  the helicity flip part of the matrix element, we find that 
it solely arises from the mass term in the expression, Eq.
(\ref{psi2}), of the
two-particle wave function . The form of the wave function  shows that this
contribution is suppressed.

The helicity dependent off-forward matrix element is conventionally
parametrized in terms of the generalized quark
distributions,
\be
\tilde F^+_{\lambda \lambda'} = {1\over { {\bar P}^+}} 
{\bar U}_{\lambda'}(P') 
\Big [\tilde  H_q({\bar x},\xi,t) \gamma^+\gamma^5 + 
\tilde E_q({\bar x},\xi,t) {\gamma^5 \Delta^+ \over {2M}} 
\Big ] U_\lambda(P), 
\label{para}
\e
where $U_\lambda(P)$ is the quark spinor in our case. The light-front spinors for
longitudinally polarized quarks are given in the appendix.
Using Eq. (\ref{a3}), and also the fact that the linear mass-dependent helicity flip
terms give suppressed contribution to the matrix element, we obtain that,
$\tilde E$ is suppressed (it has no logarithmic divergent part), provided 
$\Delta^\perp $ is small. We therefore get,
\be
\tilde H({\bar x},\xi,t)= {1\over 2} \Big [ \delta(1-{\bar x})+
{\alpha_s\over {2\pi}}C_f log{Q^2\over \mu^2}\Big ( {3\over 2} \delta(1-{\bar x}) +
{(1+{\bar x}^2-2\xi^2)\over {(1-{\bar x})_+(1-\xi^2)}}\Big ) \Big ]. 
\label{hh}
\e
The forward limit is easily obtained by putting $\xi=0$:
\be
\tilde H({\bar x},0,0) = {1\over 2} \Big [ \delta(1-{\bar x})+
{\alpha_s\over {2\pi}}C_f log{Q^2\over \mu^2}\Big ( {3\over 2} \delta(1-{\bar x}) +
{(1+{\bar x}^2)\over {(1-{\bar x})_+}}\Big ) \Big ]. 
\e
The above expression can be identified with $g_1(x)$ for a dressed quark
target, as calculated in \cite{rajen}. This gives the intrinsic helicity distribution
for a quark dressed with a gluon in perturbation theory. 
   
\subsection{Gluon Distribution}
In this section, we calculate the gluon distribution,
\be
\tilde F_{g \lambda' \lambda}^+=-{i\over {4 \pi {\bar x} {\bar P}^+}}
\int dz^- e^{{i\over 2}{\bar P}^+ { z^-} {\bar x}}\langle P'
\lambda' \mid F^{+ \alpha}(-{ z^-\over 2})\tilde  F_\alpha^+ ({ z^-\over 2}) 
\mid P \lambda \rangle.
\label{gg}
\e 
where
\be
\tilde F^{\mu \nu}={1\over 2} \epsilon^{\mu \nu \rho \sigma} F_{\rho \sigma},
~~~~~~~~~~~~\epsilon^{+1-2}=2.
\e
We use light-front gauge $A^+=0$.

The Fock space expansion of the relevant part of the operator is given by,
\be
O_g={4i\over (2 (2 \pi)^3)^2} 
\sum_\lambda \lambda \int dk_1^+ d^2k_1^\perp \int dk_2^+ d^2k_2^\perp a^\dagger
(k_1,\lambda) a(k_2,\lambda) \delta(2 {\bar x}{\bar P}^+-k_1^+-k_2^+).
\e
Here, $\lambda$ is the gluon helicity.
We calculate the matrix element for a quark state dressed with a gluon. The
Fock space expansion of the state is given by Eq. (\ref{eq2}). The single
particle sector does not contribute to the matrix element and the only contribution
comes only from the two particle sector.

The matrix element is given by,
\be 
\tilde F_g^+ = {1\over {\bar x}}\sum_\lambda \lambda  \int d^2 q^\perp \psi_2^*({1-{\bar x}\over 
{1-\xi}},q^\perp) \psi_2({1-{\bar x}\over {1+\xi}}, q^\perp+{1-{\bar x}\over 
(1-\xi^2)}\Delta^\perp) \sqrt {{\bar x}^2-\xi^2}.
\label{matrixg}
\e
We have suppressed the quark helicity dependence of the wave functions and
the sum over them.
Using the full form of the two particle wave function, 
we find that the helicity flip terms proportional to the quark mass 
give suppressed contribution and the helicity non-flip part is given by,
\be
\tilde F_g^+={\sqrt{1-\xi^2}\over {\bar x}} {\alpha_s\over 2 \pi}C_f 
log{Q^2\over \mu^2} \Big [1-{(1-{\bar x})^2\over { (1-\xi^2)}}\Big ].
\e
where the first (second) term in the square bracket comes from the state with
gluon helicity $+1(-1)$. So we have,
\be
\tilde F_g^+={\alpha_s\over 2 \pi}C_f log{Q^2\over \mu^2} {(1-(1-{\bar
x})^2-\xi^2)\over x {\sqrt {1-\xi^2}}}.
\e
Using the parametrization of $\tilde F_g^+$ in terms of $\tilde H_g$ and
$\tilde E_g$, one can write,
\be
\tilde  F^+_{g\lambda' \lambda} = {1\over { {\bar P}^+}} 
{\bar U}_{\lambda'}(P') 
\Big [ \tilde H_g({\bar x},\xi,t) \gamma^+\gamma^5 + \tilde E_g({\bar x},\xi,t) {\gamma^5
\Delta^+\over {2M}} \Big ] U_\lambda(P).
\label{parag}
\e
The fact that the helicity flip part of the matrix element is suppressed 
means that  
$\tilde E_g$ is also suppressed. 
So we get,
\be
\tilde H_g({\bar x},\xi,t)={\alpha_s\over 2 \pi}C_f log{Q^2\over \mu^2} 
{(1-(1-{\bar
x})^2-\xi^2)\over {x (1-\xi^2)}}.
\label{hg}
\e
The splitting function can easily be extracted and is given by,
\be
\tilde P_{qg}= C_f {[1-(1-{\bar x})^2-\xi^2]\over {{\bar x} (1-\xi^2)}},
\e
which again agrees with \cite{ji} when making the replacement of  $\xi$ 
of \cite{ji} by $2 \xi$. Also, in
the forward limit, Eq. (\ref{hg}) gives, 
\be
\tilde H_g( {\bar x},0,0)={\alpha_s\over 2 \pi}C_f log{Q^2\over \mu^2} 
{(1-(1-{\bar x})^2)\over {\bar x}}.
\e
or,
\be
\tilde H_g( 1-{\bar x},0,0)={\alpha_s\over 2 \pi}C_f log{Q^2\over \mu^2} 
(1+{\bar x}).
\label{intr}
\e
This gives the gluon intrinsic helicity distribution for a dressed quark
target. In Eq. (\ref{intr}), we have taken $1-{\bar x}$ as the momentum
fraction of the gluon, in order to compare with \cite{rajen}.

\section{Twist-three Distribution}
We now calculate the twist-three (transverse) component of the helicity
dependent off-forward distribution in perturbation theory.
The matrix element of the
transverse component is given by:
\be
\tilde F^\perp_{\lambda' \lambda}=\int {dz^-\over {8 \pi}} e^{{i\over 2}
{\bar P}^+{ z}^- {\bar x}}\langle P'
\lambda' \mid \bar \psi (-{z^-\over 2}) \gamma^\perp \gamma^5 \psi ({z^-\over 2}) 
\mid P \lambda \rangle.
\label{eqt}
\e

We calculate the above matrix element for a transversely polarized dressed
quark state. As before, we work in light-front gauge $A^+=0$.
The bilocal operator in this case can be written as,
\be
O^{\perp 5}=\bar \psi (-{z^-\over 2}) \gamma^\perp \gamma^5\psi ({z^-\over 2})
= \psi^{+ \dagger}(-{z^-\over 2}) \alpha^\perp \gamma^5 \psi^{- }({z^-\over 2})
+\psi^{-\dagger}(-{z^-\over 2}) \alpha^\perp \gamma^5 \psi^{+ }({z^-\over 2}).
\label{hc}
\e
The operator involves the constrained field $\psi^{- }({ z^-\over 2})$ and
therefore it is called higher twist. In the light-front gauge, $\psi^{-}$ 
can be eliminated using the constraint equation,
\be
 \psi^- = {1\over { i \partial^+}} \big [ \alpha^\perp \cdot 
( i \partial^\perp + g
A^\perp) + \gamma^0 m \big ] \psi^+,  
\e
where the operator ${1\over \partial^+}$ is defined as, using antisymmetric
boundary condition,
\be
{1\over \partial^+}f(x^-)= {1\over 4} \int_{-\infty}^{\infty} dy^-
\epsilon(x^--y^-) f(y^-).
\e
The operator, in terms of the dynamical fields, can be written as,
\be
O^{\perp 5}=O^\perp_m+O^\perp_{k^\perp}+O^\perp_g,
\e
where,
\be
O^\perp_m= m\Phi^\dagger {\sigma^1\over i \partial^+} \Phi 
+m\Big ({-\sigma^1\over i \partial^+} \Phi^\dagger\Big )\Phi
\e
\be
O^\perp_{k^\perp}=\Phi^\dagger(-{{z}^-\over 2})(-\partial^2+i\sigma_3\partial^1)
{1\over {i\partial^+}} \Phi({{z}^-\over 2})+
\Big [(\partial^2+i\sigma_3\partial^1){1\over {i\partial^+}}\Phi^\dagger
(-{{z}^-\over 2})\Big ] \Phi({{z}^-\over 2}).
\e
\be
O^\perp_g=g \Phi^\dagger(-{{z}^-\over 2}) {1\over i\partial^+}(iA^2+\sigma_3
A^1)\Phi({{z}^-\over 2})+g \Big [ {1\over {-i \partial^+}} 
 \Phi^\dagger(-{{z}^-\over 2})(-iA^2+\sigma_3 A^1)\Big ]
\Phi({{z}^-\over 2}).
\e
Here, $\Phi$ is the two component fermion field,
\be
 \psi^+ = 
\left [ \begin{array}{c} \Phi \\
                       0 \end{array} \right ].
\e
The Fock space expansion of $\Phi$ is given by Eq. (\ref{psi}), 
with $\chi_\lambda$
being the two-component spinor. The operator has three parts : $O^\perp_m$ is the
quark mass contribution, $O^\perp_{k^\perp}$ is the quark transverse momentum
contribution and $ O^\perp_g$ is the quark-gluon interaction effect. The
light-front expression clearly shows each contribution separately in
light-front gauge.

The longitudinally polarized dressed quark state is given in Eq. (\ref{eq2}). The
transversely polarized state is expressed in terms of the helicity states
as,
\be
\mid k^+, k^\perp, s^1 \rangle = {1\over {\sqrt 2}}(\mid k^+, k^\perp,
\uparrow \rangle \pm \mid k^+, k^\perp, \downarrow \rangle),
\label{c8dr1}
\e 
with $s^1 = \pm m_R$, where $m_R$ is the renormalized mass of the quark.
Without any loss of generality, we take the state to be polarized along the
$x$ direction.

The contributions to the matrix element coming from the three parts of the
operator are given by,
\be
\tilde F^1_m &=& {m\over {\bar P}^+} \Big [ 
\delta(1-{\bar x})\psi_1^*\psi_1
\nonumber \\&&~~~~~+\sum_{\sigma,\sigma'}\int d^2
q^\perp {{\bar x}\over {{\bar x}^2-\xi^2}} \psi_2^*({{\bar x}-\xi\over {1-\xi}},
q^\perp+{1-{\bar
x}\over {1-\xi^2}}\Delta^\perp)\chi^\dagger_\sigma \sigma^1
 \chi_{\sigma'} \psi_2({{\bar x}+\xi\over {1+\xi}}, q^\perp)\Big ].
\e
We have suppressed the  quark helicity dependence of the wave function.
Using the explicit form of the two particle wave function, 
\be
\tilde F^1_m = {m\over {\bar P}^+} {1\over \sqrt {1-\xi^2}}\psi_1^* \psi_1
\Big [ \delta(1-{\bar x})
+{\alpha_s\over {2
\pi}}C_f log{Q^2\over \mu^2}\Big (
{2 {\bar x} ({\bar x}-2\xi^2)\over {(1-{\bar x})({\bar x}^2-\xi^2)}}\Big ) 
\Big ],
\label{mm}
\e
\be
\tilde F^1_{k^\perp}&=& - i\sum_{\sigma,\sigma'} \int d^2 q^\perp \psi_2^*({{\bar x}-
\xi\over {1-\xi}},q^\perp+{1-{\bar
x}\over {1-\xi^2}}\Delta^\perp) \psi_2({{\bar x}+\xi\over {1+\xi}}, q^\perp)
{q^2\over {\bar P}^+} {\xi \over {{\bar x}^2-\xi^2}}+\nonumber\\&&~~~~+
 \sum_{\sigma,\sigma'} \int d^2 q^\perp \psi_2^*({{\bar x}-\xi\over {1-\xi}},
q^\perp+{1-{\bar
x}\over {1-\xi^2}}\Delta^\perp)\chi^{\dagger}_{\sigma} {( \sigma^3 q^1 )
\over {\bar P}^+} \chi_{\sigma'}  \psi_2({{\bar x}+\xi\over {1+\xi}}, q^\perp) 
   {{\bar x}  \over {{\bar x}^2-\xi^2}},
\e
This  gives,
\be
\tilde F^1_{k^\perp}=-{m\over {\bar P}^+}  {1\over \sqrt {1-\xi^2}}
 C_f log{Q^2\over \mu^2}{\alpha_s\over 
{2 \pi}}{{(1-\bar x}) ({\bar x}^2+\xi^2+2 {\bar x} \xi^2)\over {({\bar x}^2-\xi^2)(1-\xi^2)}}.
\label{mk}
\e
The interaction part gives overlap contribution in terms of two-and one
particle wave function and is given by,
\be
\tilde F^1_g=C_f log{Q^2\over \mu^2}{\alpha_s\over 
{2 \pi}}{m\over {2 \bar P}^+}{1\over \sqrt {1-\xi^2}} \delta(1-{\bar x})
\label{intg}
\e
As before, we have taken $\Delta^\perp$ to be small. The interaction gives
contribution only at the end point ${\bar x}=1$. Considering the
normalization contribution to the single particle matrix element, we get,
the total contribution,
\be
\tilde F^1 = {m\over {\bar P}^+} {1\over \sqrt {1-\xi^2}}\Big [  
 \delta(1-{\bar x})+  C_f log{Q^2\over \mu^2}{\alpha_s\over 
{2 \pi}} \Big ( 2 \delta(1-{\bar x}) +
{1+2{\bar x}(1-\xi^2)-{\bar x}^2\over {(1-{\bar x})_+
(1-\xi^2)}} \Big ) \Big ],
\e
Here, we have also considered the contribution of the normalization
condition to the single particle matrix element, which cancels the end point
singularity, similar to the twist-two case. Here,
$m$ is
the bare quark mass. The above expression has no singularity at ${\bar
x}=\xi$. In light-front theory, the linear mass term appearing
in the light-front QCD Hamiltonian is renormalized as \cite{perry},
\be
m_R=m \Big ( 1+{3\over {4\pi}} \alpha_s C_f log{Q^2\over \mu^2} \Big ).
\label{renorm}
\e
Here $m_R$ is the renormalized mass of the quark. The linear mass terms in
light-front QCD Hamiltonian are associated with explicit chiral symmetry
breaking \cite{wil}. Also, from the above results, we find that the three
contributions, including the quark transverse momentum effect and the quark
gluon interaction effect are proportional to the quark mass, which shows
that the twist-three distribution is directly related to the dynamical
effect of  chiral symmetry breaking. In terms of the renormalized mass, we
get,
\be
\tilde F^1 = {m_R\over {\bar P}^+} {1\over \sqrt {1-\xi^2}}\Big [  
 \delta(1-{\bar x})+  C_f log{Q^2\over \mu^2}{\alpha_s\over 
{2 \pi}} \Big ( {1\over 2} \delta(1-{\bar x}) +{1+2{\bar x}(1-\xi^2)-
{\bar x}^2\over {(1-{\bar x})_+ (1-\xi^2)}} \Big ) \Big ],
\label{tildef}
\e
In the forward limit, this gives,
\be
\tilde F^1 = {m_R\over {\bar P}^+} \Big [  
 \delta(1-{\bar x})+  C_f log{Q^2\over \mu^2}{\alpha_s\over 
{2 \pi}} \Big ( {1\over 2} \delta(1-{\bar x}) + {(1+2 {\bar x}-{\bar x}^2)
\over {(1-{\bar x})_+}}\Big ) \Big ].
\e
By comparing the {\it rhs} of the above equation with  
the transversely polarized structure function $g_T$ for a 
dressed quark target \cite{gt}, one obtains that,
\be
\tilde F^1= {2m_R\over {\bar P}^+} g_T ={2S_T\over {\bar P}^+} g_T ,
\e
 Since for a transversely polarized dressed quark, $m_R=S_T$ (see appendix).
\section{Examination of the Wandzura-Wilczek Relation in Perturbation
Theory}

The twist-three matrix element is parametrized as \cite{kiptily},
\be
\tilde F^{\perp }&=& {1\over {\bar P}^+} {\bar U (P')} \Big (
\gamma^\perp \gamma^5 {\tilde H} + 
{{\Delta^\perp} \over {2 M}} \gamma^5 \tilde E+
{\Delta^\perp \gamma^5\over 2 M} \tilde G_1+ \gamma^\perp
\gamma^5 {\tilde G_2} 
+\Delta^\perp \frac{\gamma^+}{\bar P^+} \gamma^5 \tilde G_3 \nonumber
\\&& ~~~~~~~~~~~~~~~~~~~~~~~+i \epsilon^{\perp \nu} 
\Delta_{\nu} \frac{\gamma^+ }{\bar P^+}\tilde G_4 \Big ) U(P).
\label{t3}
\e
The light-front spinors for a transversely polarized quark are given by
Eq. (\ref{spintran}) in the appendix.
Using Eq. (\ref{t3}),(\ref{spi}),(\ref{spinor}) we get,
\be
\tilde F^1 = 
{2M\over {\sqrt {1-\xi^2} {\bar P}^+}} (\tilde H+\tilde G_2),
\label{ftil1}
\e
which in the forward limit gives ${2S_T\over {\bar P}^+}g_T$, since $\tilde
H(x,0,0)=g_1(x)$ and $\tilde G_2(x,0,0)=g_2(x)$ in the forward limit.
Comparing with the result in the previous section, we see that Eq. 
(\ref{ftil1}) is in
agreement with our result for a dressed quark.

Using Eq. (\ref{hh}) and (\ref{tildef}) we get,
\be
\tilde G_2({\bar x},\xi,t)= {1\over 2} C_f log{Q^2\over \mu^2}{\alpha_s\over {2 \pi}}
\Big [ -\delta(1-{\bar x})+{2({\bar x}+\xi^2)\over {(1-\xi^2)}}\Big ],
\e
which in the forward limit gives,
\be
\tilde G_2({\bar x},0,0)= {1\over 2}C_f log{Q^2\over \mu^2}{\alpha_s\over {2 \pi}}
\Big [ -\delta(1-{\bar x})+2{\bar x}\Big ].
\e
The above expression agrees with $g_2$ for a transversely polarized dressed
quark target \cite{gt}.  

In the Wandzura-Wilczek approximation, where the quark mass as well as the
quark-gluon interaction terms are neglected, the twist-three matrix element is
given in terms of twist-two matrix elements as \cite{wan},
\be
\tilde F_\mu^{ WW}(x,\xi)&=&{\bar U}(P') \Big [ {\Delta^\mu \gamma^5
\over {2M}} \tilde
E(x,\xi)-{\Delta^\mu\over {2\xi}} \gamma^+\gamma^5 \tilde H(x,\xi)\Big ]
U(P)\nonumber\\
&&~~~~+\int_{-1}^1 du {\tilde G}_\mu (u,\xi)W_+(x,u,\xi)+
i\epsilon_{\perp \mu k} 
 \int_{-1}^1 du { G}^k (u,\xi)W_-(x,u,\xi),
\e
where,
\be
G^\mu (u,\xi)&=&{\bar U}(P') \Big [ \gamma^\mu_\perp (H+E)(u,\xi)+
{\Delta^\mu\over {2\xi}} {1\over M} \Big ( u{\partial \over {\partial u}}+\xi
{\partial \over {\partial \xi}}\Big ) E(u,\xi)\nonumber\\&&~~
- {\Delta^\mu\over {2\xi}} \gamma^+ \Big ( u{\partial \over {\partial u}}+\xi
{\partial \over {\partial \xi}}\Big ) (H+E)(u,\xi)\Big ] U(P),
\e
\be
\tilde G^\mu (u,\xi)&=&{\bar U}(P') \Big [ \gamma^\mu_\perp \gamma^5 
\tilde H(u,\xi)+
{\Delta^\mu\over {2}} {\gamma_5\over M} \Big (1+ u{\partial \over {\partial u}}
+\xi {\partial \over {\partial \xi}}\Big ) \tilde E(u,\xi)\nonumber\\&&~~
- {\Delta^\mu\over {2\xi}} \gamma^+ \gamma_5\Big ( u{\partial \over {\partial u}}+\xi
{\partial \over {\partial \xi}}\Big ) \tilde H (u,\xi)\Big ] U(P),
\e
$W_{\pm}(x,u,\xi)$ are the Wandzura-Wilczek kernels given by,
\be
W_{\pm}(x,u,\xi)&=& \Big [ \theta(x > \xi){\theta(u > x)\over
(u-\xi)} -\theta(x < \xi){\theta(u < x)\over (u-\xi)} \Big ]\nonumber\\
&&\pm \Big [ \theta(x > -\xi){\theta(u > x)\over (u+\xi)}
- \theta(x < -\xi){\theta(u < x)\over (u+\xi)} \Big ].
\e
Using the light-front spinors given in the appendix, we find, for $x>\xi$ and
$\Delta^\perp=\Delta^1; \Delta^2=0$ that the WW
relation for $ \tilde F^{\perp}_{WW}$ reduces to,
\be
\tilde F^1_{WW}={2 m_R\over {{\bar P}^+ \sqrt {1-\xi^2}}} \int du {\tilde H}
(u,\xi) {\theta(x-\xi) \theta(u-x)\over {u-\xi}}.
\e
In the forward limit, the {\it rhs } becomes ${2 m_R\over P^+} \int_x^1 dy
{g_1(y)\over y}$, which gives the well known Wandzura-Wilczek relation for the
transversely polarized DIS structure function $g_T$. The twist three vector
distribution $F^\perp $ is similarly expressed in terms of a Wandzura-Wilczek
relation, however it vanishes in the forward limit. 
Using the expression for $\tilde H $ for a massive dressed quark in
perturbation theory, 
\be
\tilde F^1_{WW}&=& {2 m_R\over {{\bar P}^+ \sqrt {1-\xi^2}}} {1\over 2}
\theta({\bar x}-\xi) \Big [ {\theta(1-\xi)\over (1-\xi)}+{\alpha_s\over 2 \pi}
C_f log{Q^2\over \mu^2}\Big [ {3\over 2} {\theta (1-\xi)\over
1-\xi}\nonumber\\&&~~~~~~~~~~~~
+{1\over (1-\xi^2)} \Big ( (1+\xi)log\Big ({1-\xi\over {\bar x}-\xi}\Big )-1
+{\bar x}\Big )\Big ].
\e
Comparing the above expression with Eq. (\ref{tildef}), 
we see that the WW relation is not satisfied for a dressed quark state in
perturbation theory, similar to the forward case.  This is not surprising because in the WW approximation,
the mass of the quark as well as the explicit interaction dependence of the
operator is neglected, whereas,  we have obtained the full result in
perturbative QCD for a massive quark.The
effect of quark transverse momentum in $g_T$ was investigated in a covariant
parton model approach in \cite{cov}. Also, it is known that in the forward
limit, the WW relation is violated in perturbation theory \cite{gt}.  However,
BC sum rule is satisfied \cite{alt,gt,bur}.

The quark mass effect can be incorporated in the derivation of the
off-forward WW relation \cite{kiptily}. This gives an additional contribution to $\tilde F^\perp$ which is of the form 
(see Appendix B),
\be
\tilde F^\perp_{mass}={2 m \over {\bar P^+}}\Big [ -{{\bar x}\over {{\bar x}^2-\xi^2}} f^\perp({\bar x},\xi,\Delta) +\int_{\bar x}^1 dy {y^2+\xi^2\over (y^2-\xi^2)^2}f^\perp (y,\xi,\Delta)\Big ],
\label{kip}
\e
for $\xi <{\bar x}<1$, where
\be
f^\perp({\bar x},\xi,\Delta)={1\over 2}\int {dz^-\over 2 \pi}e^{-{i\over 2}
 {\bar P}^+ z^- {\bar x}}
\langle P'
\lambda' \mid \bar \psi (-{z^-\over 2})i \sigma^{+ \perp } \gamma^5 \psi ({z^-\over 2}) 
\mid P \lambda \rangle.
\e
We use the parametrization \cite{diehlm},
\be
{1\over 2}\int {dz^-\over 4 \pi}e^{-{i\over 2} {\bar P}^+ z^-{\bar x}}
\langle P'
\lambda' \mid && \bar \psi (-{z^-\over 2})i \sigma^{+j } \gamma^5 \psi ({z^-\over 2}) 
\mid P \lambda \rangle \nonumber\\&&={1\over {\bar P}^+}{\bar U}(P',\lambda')\Big [ H_T^q i\sigma^{+ j} \gamma^5
+\tilde H_T^q  {i\epsilon^{+ j \alpha \beta}\Delta_\alpha P_\beta \over M^2}
+E_T^q
i{\epsilon^{+ j \alpha \beta}\Delta_\alpha \gamma_\beta \over 2
M}\nonumber\\&&~~~~~~~~~~~~~~~~~~~~+\tilde E_T^q 
{i\epsilon^{+ j \alpha \beta}P_\alpha \gamma_\beta \over M}\Big ] U(P,\lambda).
\e
Here $M$ is the mass of the state.
We calculate the above matrix element for a transversely polarized dressed quark state in perturbation theory. 
Using the relations of light cone spinors, and also using the normalization
of the transversely polarized state, we obtain,
\be
H_T^q={1\over 2} \Big [\delta(1-{\bar x})+  C_f log{Q^2\over \mu^2}{\alpha_s\over 
{2 \pi}} \Big ( {3\over 2} \delta(1-{\bar x}) + {2 ( {\bar x}-\xi^2)
\over {(1-{\bar x})_+(1-\xi^2)}}\Big ) \Big ],
\e
which in the forward limit gives $h_1(x)$ for a dressed quark :
\be
h_1({\bar x})={1\over 2} \Big [\delta(1-{\bar x})+  C_f log{Q^2\over \mu^2}{\alpha_s\over 
{2 \pi}} \Big ( {3\over 2} \delta(1-{\bar x}) + {2 {\bar x}
\over {(1-{\bar x})_+}}\Big ) \Big ].
\e
Next, we investigate the mass corrections to the WW relation and the `genuine twist three contribution ' 
to the matrix element in somewhat more detail. In the forward limit, ${\tilde F}^\perp$ corresponds to $g_T$. 
We can write, in the forward limit,
\be
g_T(x)=\int_x^1 dy {g_1(y)\over y} +{m\over M}\Big ({ h_1(x)\over x } -\int_x^1 dy {h_1(y)\over y^2}
\Big ) +g_T^g(x),
\label{mass}
\e
where $g_T^g(x)$ is the so-called `genuine twist three' contribution to $g_T$. If we neglect this, we get the WW relation with the quark mass correction,
\be
g_T(x)=\int_x^1 dy {g_1(y)\over y} +{m\over M}\Big ({ h_1(x)\over x } -\int_x^1 dy {h_1(y)\over y^2}
\Big ).
\e
Here $m$ is the quark mass and $M$ is the mass of the target. It is very important to note that in perturbative calculation, $m$ has to be renormalized. Taking $n$-th moment of both sides of Eq. (\ref{mass}) we get,
\be
g_2^n=-{n\over n+1} g_1^n+{m\over M}{n\over n+1}h_1^{n-1}+g_{T g}^n,
\e
where $a^n=\int_0^1 dx x^n a(x)$.
Using the expressions of $g_1(x),g_2(x)$ and $h_1(x)$, the moments can be directly calculated,
\be
g_2^n={1\over 2} {\alpha_s\over 2\pi} C_f log{Q^2\over \mu^2} (-{n\over n+2}),
\e
\be
g_1^n={1\over 2}\Big [1+ {\alpha_s\over 2\pi} C_f log{Q^2\over \mu^2} 
\Big (-{1\over 2}+
{1\over {(n+1)(n+2)}}-2\sum_{j=2}^{n+1} {1\over j}\Big ) \Big ],
\e
\be
h_1^n={1\over 2}\Big [1+ {\alpha_s\over 2\pi} C_f log{Q^2\over \mu^2} \Big ({3\over 2}
-2\sum_{j=1}^{n+1} {1\over j}\Big ) \Big ].
\e
Using these and also the renormalization of quark mass given by Eq. (\ref{renorm}), we obtain,
\be
g_{T g}^n= {1\over 2}{\alpha_s\over 2\pi} C_f log{Q^2\over \mu^2}\Big [-{n\over n+2}+{n\over n+1}
\Big ( {3\over 2}-{2n+3\over {(n+1)(n+2)}}\Big )\Big ].
\label{eq:gtn}
\e
For $n=0$, the {\it rhs} of the above equation gives zero, 
which proves the BC sum rule. For $n=1$ , 
the {\it rhs} of Eq.~(\ref{eq:gtn}) also yields zero, 
which gives the Efremov-Leader-Teryaev sum rule with the correction 
due to quark mass.

Next, we use Eq. (\ref{mass}) to extract the `genuine twist three' part of $g_T$.

The $O(\alpha_s)$ part of $g_T$ can  be separated into two parts,
\be
g_T^{(1)}= g_{T A}^{(1)}+ g_{T B}^{(1)}.
\e
Here $g_{T A}^{(1)}$ is the WW part with the mass corrections and
$g_{T B}^{(1)}$ is the `genuine twist three part'. 
Using Eq. (\ref{mass}) and also the expressions of $g_1(x), h_1(x)$ we get,
\be
g_{T B}^{(1)}= {1\over 2} {\alpha_s\over 2\pi} C_f log{Q^2\over \mu^2} \Big [ {1\over 2}\delta(1-x)
-{3\over 2} - log x \Big ].
\e
It is interesting to compare the {\it rhs} of the above equation with the forward limit of Eq. (\ref{intg}).
This shows that Eq.(\ref{intg}) does not give the full `genuine' twist three contribution but only a part of it. Also from the above expression it is easy to check that the first and second moments of
 $g_{T B}^{(1)}$ are zero.

We stress that, 
the quark mass plays a very important role in the twist-three matrix
element, and also, in our case, it is essential to obtain a
transversely polarized state, since $S_T=m_R$, the renormalized mass of the
quark. Our result shows that the twist-three
generalized distribution is directly related to the chiral symmetry breaking
dynamics in light-front QCD. 

\section{Summary and Discussions}

To summarize, in this work, we have investigated the off-forward matrix
elements of the light-cone axial vector operator. We have calculated the
matrix elements of the plus and transverse components of the operator for a
dressed quark in light-front Hamiltonian perturbation theory. This approach
allows us to express the distributions in terms of light-front wave
functions. We have
restricted ourselves to the kinematical region $\xi <{\bar x}<1$.   
In this case, the overlaps of three-particle and one-particle wave functions are
absent. We obtained the splitting functions for the evolution of the helicity dependent twist
two quark and gluon distributions in a straightforward way. We showed that the
singularity at ${\bar x}=1$ is canceled by the contribution from the
normalization of the state, similar to the helicity independent case
calculated earlier. The twist-two distributions reduce to the quark and
gluon intrinsic helicity distributions for a dressed quark target in the
forward limit. The twist-three distribution is expressed entirely in terms
of the dynamical fields in the light-front gauge. 
This calculation shows that for the twist-two distributions,
the entire interaction dependence comes from the state whereas the operator
has free field structure, but in the case of twist-three, both the operator
and the state introduce interaction dependence. The operator has three
parts, explicit mass dependent term, quark-gluon interaction term and a
term containing the quark transverse momentum effect. The calculation of this matrix
element for a transversely polarized dressed quark shows that all 
the three 
contributions are proportional to the quark mass. Using the renormalized
quark mass $m_R$ in light-front Hamiltonian perturbation theory, we found that in
the forward limit, $\tilde F^\perp$ is proportional to $S_T g_T$, where
$S_T$ is the transverse polarization of the state, $S_T=m_R$ in our case. It
is known that in light-front Hamiltonian QCD, chirality is the same as
helicity, and the terms that cause helicity flip in the light-front QCD
Hamiltonian are explicit chiral symmetry breaking terms. These terms are
linear in the quark mass. It is interesting to note that the quadratic mass
terms do not flip helicity, however, they are suppressed here. Therefore, we
concluded that $\tilde F^\perp$ is directly related to the chiral symmetry
breaking dynamics in light-front QCD and the quark mass plays an important
role. In particular, a finite mass is necessary to have a transversely polarized quark
state. We have calculated the same off-forward matrix element in the
Wandzura-Wilczek approximation and found that the actual result for a massive
dressed quark deviates from the WW approximated form. The violation of the WW
relation for $g_T$ for a massive quark is known in perturbation theory and our result reduces to
$g_T$ for a massive dressed quark in the forward limit. It is to be noted
that in the case of nucleons, the
quark intrinsic transverse momentum effects and the quark-gluon
coupling dynamics play a more complicated role and the pure quark mass
effects in perturbative QCD may be suppressed by ${m\over M}$ where $M$ is
the hadron mass. We have also calculated the quark mass correction to the off forward WW relation which in the forward limit reduces to a term proportional to $h_1(x)$. We extracted the 'genuine twist three part' of $g_T$
and showed that both BC and ELT sum rules are satisfied.

It is known that in the kinematical region $0<{\bar x}<\xi$, 
contribution comes from the overlap of three-and-one-particle wave functions. 
The  GPDs in this region have a different type of evolution (Brodsky-Lepage).
 It will be interesting to investigate the GPDs for a dressed quark in this kinematical region using this approach and to check the various moment relations in the whole range of ${\bar x}$, $0<{\bar x}<1$. Another interesting topic for future work is to investigate the $\Delta^\perp$ dependence of the GPD's in the frame $\xi=0$. 
\section{acknowledgments}

We thank M. V. Polyakov for many helpful discussions during the course of this
 work. We thank D. Kiptily for helping us in the derivation of 
Eq. (\ref{kip}). AM also thanks E. Reya and A. Harindranath for helpful 
discussions and comments. 
This work was supported by a grant of the 
Zentrum f\"ur physikalisch-chemische Verbundforschung at the Universit\"at
Mainz and by the Deutsche Forschungsgemeinschaft (SFB 443). At
the Universit\"at Dortmund, this work has been supported in part by the
Bundesministerium f\"ur Bildung und Forschung, Berlin/Bonn.

\appendix
\section{Light-front spinors}
The light-front spinors for longitudinally polarized quark of mass $M$ and
momentum $P$ and helicity up and down, respectively, are given by \cite{ped},

\be
U_{\uparrow} (P) = {1\over \sqrt {2 P^+}}\left (\begin{array}{c}P^++M\\P^1+iP^2\\
P^+-M\\P^1+iP^2\end{array}\right),~~~~~~~~~~~~~~~~
U_{\downarrow} (P) = {1\over \sqrt {2P^+}} 
\left (\begin{array}{c}-P^1+iP^2\\P^++M\\
P^1-iP^2\\-P^++M\end{array}\right).
\e
Using these, we get,
\be
{\bar U}_{\uparrow}(P')\gamma^+ \gamma^5  U_{\uparrow}(P) =2 {\sqrt
{1-\xi^2}} {\bar P}^+\nonumber\\
{\bar U}_{\uparrow}(P') \gamma^5  U_{\uparrow}(P) = { 2 \xi M \over {\sqrt
{1-\xi^2}} }.
\e
Also,
\be
{\bar U}_{\uparrow}(P')\gamma^+ \gamma^5  U_{\downarrow}(P) =0
\nonumber\\
{\bar U}_{\uparrow}(P') \gamma^5  U_{\downarrow}(P) = { -\Delta^1+i
\Delta^2 \over {\sqrt {1-\xi^2}} }.
\label{a3}
\e
Light-front spinors for transversely polarized quark are given by
\cite{thesis},
\be
U_{\uparrow} (P) = {1\over \sqrt {2P^+}}\left (\begin{array}{c}M+P^+-iP^2\\
-P^1\\
P^1\\-M+P^++iP^2\end{array}\right),~
U_{\downarrow} (P) = {1\over \sqrt {2P^+}} 
\left (\begin{array}{c}-M+P^+-iP^2\\-P^1\\
P^1\\M+P^++iP^2\end{array}\right).
\label{spintran}
\e
Using these, we get the components of the polarization vector $S^\mu={1\over
2}{\bar U} (P) \gamma^\mu \gamma^5 U(p)$ : $S^+=0$, 
 $S^2=0$, $S^1=M$, $ S^-=2 {P^1 \over P^+}M$.

Also,
\be
{\bar U}_{\uparrow}(P')\gamma^\perp \gamma^5  U_{\uparrow}(P) ={2 M\over 
{\sqrt {1-\xi^2}}} \nonumber\\
{\bar U}_{\uparrow}(P') \gamma^5  U_{\uparrow}(P) = 0.
\label{spi}
\e
Also,
\be
{\bar U}_{\uparrow}(P')\gamma^+ \gamma^5  U_{\uparrow}(P) =0,
\nonumber\\
{\bar U}_{\uparrow}(P') \gamma^1  U_{\uparrow}(P) = { \xi
\Delta^1 \over {\sqrt {1-\xi^2}}} \nonumber\\
{\bar U}_{\uparrow}(P') U_{\uparrow}(P) ={2 M\over 
{\sqrt {1-\xi^2}}} \nonumber\\
{\bar U}_{\uparrow}(P') \gamma^+  U_{\uparrow}(P) = 2 {\bar P}^+ {\sqrt
{1-\xi^2}}.
\label{spinor}
\e

\section{Mass Term in WW relation}

In this appendix, we give an outline of the derivation of Eq.
(\ref{kip}). Using the approach described in \cite{kiptily}, and taking
into account the quark mass $m$, one gets,
\be
\tilde F^\alpha (x,\xi,t)&=&\int {d \lambda\over 2 \pi} e^{-i \lambda x}  \langle P'
S' \mid \bar \psi (-{z^-\over 2}) \gamma^\alpha \gamma^5 \psi ({z^-\over 2}) 
\mid P S \rangle \nonumber\\&&~~~~~~~~~~~~=  M^\alpha(x,\xi,t)+X^\alpha(x,\xi,t),
\e
were $\lambda={1\over 2} {\bar P^+} z^- $, $M^\alpha (x,\xi,t)$ is the mass term and $X^\alpha(x,\xi,t)$ are all
the other terms considered in the WW approximation. Here we concentrate on
the mass term given by \cite{maxim},

\be
M^\alpha(x,\xi,t)&=&-im \int {d\lambda\over 2 \pi} e^{-i \lambda x}\lambda \int_0^1 du u \Big [ e^{i (1-u) \xi
\lambda}+e^{-i (1-u) \xi \lambda}\Big ]\nonumber\\&&~~~~~~~~~~~~~~
{1\over 2} \langle P'
S' \mid \bar \psi (-{uz^-\over 2}) i \sigma^{+\alpha}\gamma^5 \psi ({uz^-\over 2}) 
\mid P S \rangle. 
\label{malpha}
\e
 If we define,
\be
f^\alpha(x,\xi,\Delta)={1\over 2}\int {dz^-\over 2 \pi}e^{-{i\over 2}
 {\bar P}^+ z^-  x}
\langle P'
S' \mid \bar \psi (-{z^-\over 2})i \sigma^{+ \alpha } \gamma^5 \psi ({z^-\over 2}) 
\mid P S \rangle,
\e
we get, from Eq. (\ref{malpha}),
\be
M^\alpha(x,\xi,t)={m\over {\bar P^+}}{\partial\over {\partial x}} \int_0^1 du \Big
[f^\alpha({x+(1-u)\xi \over u},\xi,\Delta) +f^\alpha({x-(1-u)\xi\over u},\xi,\Delta)
\Big ].   
\e
Changing the variable, $y={x\pm (1-u) \xi\over u}$  we obtain, for  $x>\xi$,
\be
M^\alpha(x,\xi,t)={2 m \over {\bar P^+}}\Big [ -{x\over {{\bar x}^2-\xi^2}} 
f^\alpha(x,\xi,\Delta) +\int_x^1 dy {y^2+\xi^2\over
(y^2-\xi^2)^2}f^\alpha (y,\xi,\Delta)\Big ].
\e


\begin{thebibliography}{99}

\bibitem{dieter} D. M\"uller, D. Robaschik, B. Geyer, F. M. Dittes, J.
Horejsi, Fortsch. Phys. {\bf 42}, 101 (1994).

\bibitem{rev1} X. Ji, J. Phys. {\bf G 24}, 1181 (1998).

\bibitem{rev2}  A. V. Radyushkin,
hep-ph/0101225, published in "At the Frontier of Particle Physics/Handbook
of QCD", ed. M. Shifman (World Scientific, Singapore, 2001).

\bibitem{rev3}  K. Goeke, M.
V. Polyakov, M. Vanderhaeghen, Prog. Part. Nucl. Phys. {\bf 47}, 401 (2001). 

\bibitem{brod} S. J. Brodsky, M. Diehl, D. S. Hwang, Nucl. Phys. {\bf B 596},
99 (2001).


\bibitem{kroll} M. Diehl, T. Feldmann, R. Jakob, P. Kroll, Nucl. Phys. {\bf
B 596}, 33 (2001).

\bibitem{miller} M. Diehl, T. Feldmann, R. Jakob, P. Kroll, Eur. Phys. J {\bf C8}, 409 (1999);
B. C. Tiburzi and G. A. Miller, Phys. Rev. {\bf C 64},
065204 (2001); 
A. Mukherjee, I. V. Musatov, H. C. Pauli, A. V. Radyushkin, hep-ph/0205315. 

\bibitem{hermes} A. Airapetian {\it et al}, (HERMES Collaboration), Phys.
Rev. Lett. {\bf 87}, 182001 (2001).

\bibitem{clas} S. Stepanyan {\it et al}, (CLAS Collaboration), Phys. Rev.
Lett. {\bf 87}, 182002 (2001).

\bibitem{h1} C. Adloff {\it et al}, (H1 Collaboration), Phys. Lett. {\bf B
517}, 47 (2001).

\bibitem{zeus} P. R. Saull {\it et al}, (ZEUS Collaboration),
hep-ex/0003030.

\bibitem{wan} A. V. Belitsky and D. M\"uller, Nucl. Phys. {\bf B589}, 611
(2000); N. Kivel, M. V. Polyakov, A. Sch\"afer and O. V. Teryaev, Phys.
Lett. {\bf B497},73 (2001).

\bibitem{ww} S. Wandzura and F. Wilczek, Phys. Lett. {\bf B72}, 195 (1977).

\bibitem{exp} P. L. Anthony {\it et al}, (E155 Collaboration), hep-ex/0204028.
 
\bibitem{hari1} W. M. Zhang and A. Harindranath, Phys. Rev. {\bf D48}, 4881
(1993).

\bibitem{hari2} A. Harindranath, R. Kundu, W. M. Zhang, Phys. Rev 
{\bf D 59}, 094013 (1999). 

\bibitem{rajen} A. Harindranath and R. Kundu, Phys. Rev. {\bf D 59}, 116013
(1999).

\bibitem{gt} A. Harindranath and W. M. Zhang, Phys. Lett. {\bf B408}, 347
(1997).

\bibitem{hari3} A. Harindranath, A. Mukherjee, R. Ratabole, Phys. Rev. 
{\bf D 63}, 045006 (2001).

\bibitem{hari4} A. Harindranath, R. Kundu, A. Mukherjee and J. P. Vary, Phys. Lett.
{\bf B 417}, 361 (1997); Phys. Rev. {\bf D 58}, 114022 (1998). 
 
\bibitem{tran} A. Mukherjee and D. Chakrabarti, Phys. Lett. {\bf B
506}, 283 (2001).

\bibitem{metz} R. Kundu and A. Metz, Phys. Rev. {\bf D 65}, 014009 (2002).

\bibitem{vec} A. Mukherjee and M. Vanderhaeghen, Phys. Lett. {\bf B 542}, 245
(2002).

\bibitem{bc} H. Burkhardt and W. N. Cottingham, Ann. of Phys, {\bf 56}, 453 (1970).

\bibitem{elt} A. V. Efremov, O. V. Teryaev and E. Leader, Phys. Rev. {\bf D 55}, 4307 (1997).
 
\bibitem{ji} X. Ji, Phys. Rev. {\bf D55}, 7114 (1997).

\bibitem{perry} R. J. Perry, Phys. Lett. {\bf B 300}, 8 (1993); A. Harindranath
and W. M. Zhang, Phys. Rev. {\bf D 48}, 4903 (1993).

\bibitem{wil} K. G. Wilson, T. S. Walhout, A. Harindranath, W. M. Zhang, R.
J. Perry and S. D. Glazek, Phys. Rev. {\bf D 49}, 6720 (1994).


\bibitem{kiptily} D. V. Kiptily and M. V. Polyakov, Preprint, hep-ph/0212372.


\bibitem{cov} J. D. Jackson, G. G. Gross and  R. G. Roberts, Phys. Lett.
{\bf B 226}, 159 (1989).  

\bibitem{alt} G. Altarelli, B. Lampe, P. Nason and G. Ridolfi, Phys. Lett.
{\bf B 334}, 187 (1994).

\bibitem{bur} M. Burkardt and Y. Koike, Nucl. Phys. {\bf B 632}, 311 (2002).

\bibitem{maxim}D. V. Kiptily and M. V. Polyakov, private communication.

\bibitem{diehlm} M. Diehl, Eur. Phys. J. {\bf C 19}, 485 (2001).


\bibitem{ped} A. Harindranath, An Introduction to the Light Front
Dynamics for Pedestrians in {\it Light-front Quantization and
Non-perturbative QCD } , Ed. J. P. Vary and F. Wolz, published by
Internatinal Institute of Theoretical and Applied Physics, Ames, Iowa, USA
(1997).

\bibitem{thesis} A. Mukherjee, Ph. D. Thesis, University of Calcutta,
hep-ph/0106167.

\end{thebibliography}
\end{document}